# Participant-Mediated Collection of Sensitive Digital Trace Data: The CANDOR Research Infrastructure

**Andrew Zhao**[1*] · **Rijul Magu**[2] · **Ekta Raj**[3] · **Teresa Elinjikkal**[3] · **Munmun De Choudhury**[2*†]

* These authors contributed equally to this work.

† Corresponding author: Munmun De Choudhury (munmund@gatech.edu).

[1] Institute for People and Technology, Georgia Institute of Technology, Atlanta, GA, USA

[2] School of Interactive Computing, Georgia Institute of Technology, Atlanta, GA, USA

[3] College of Computing, Georgia Institute of Technology, Atlanta, GA, USA

## Abstract

Digital trace data provide rich measures of behavior in everyday settings, but the research ecosystem supporting their collection has become increasingly constrained by declining platform API access and a historical reliance on publicly observable data. Participant-mediated data donation offers a complementary approach in which individuals contribute selected portions of their own digital histories to research. Such data can include longitudinal and non-public behavior, span multiple platforms and modalities, and be linked to independently collected study measures, enabling study designs that are difficult to implement using public social media data alone. These opportunities also introduce methodological challenges around participant control, data minimization, heterogeneous platform exports, privacy, and governance, particularly when semantic or multimodal content is necessary to study the construct of interest. We present CANDOR (Collecting and Analyzing Networked Data for Open Research), an end-to-end infrastructure for participant-mediated collection and governance of sensitive digital trace data. CANDOR supports participant-directed selection of platforms, data types, and temporal ranges; modular platform- and modality-specific parsing and de-identification; linkage to independent study measures; and protected processing, storage, and access. We derive design requirements for this class of research and compare CANDOR with existing data donation infrastructures, identifying how different approaches support participant control, data minimization, scientifically necessary data richness, study-design flexibility, and governance. Together, this work provides a methodological and infrastructural framework for using participant-contributed digital traces in behavioral research, particularly when the data needed to address a scientific question are longitudinal, non-public, multimodal, or sensitive.

*Keywords*: data donation; digital trace data; participant-mediated research; research infrastructure; privacy; social media; multimodal data

## 1. Introduction

Digital traces from social media and other online platforms have become an important source of evidence about behavior as it unfolds in everyday life. Over the past decade and a half, computational social science has used these data to study social interaction, information exposure, health and well-being, collective behavior, and other phenomena at a scale and temporal resolution that are difficult to obtain through conventional surveys or laboratory studies (Lazer et al., 2009, 2020; Lazer & Radford, 2017; Salganik, 2018). Yet this research tradition has also been shaped by the data that happened to be accessible. Twitter, in particular, became what Tufekci (2014) described as a “model organism” for social media research, not because it represented the full range of online social behavior, but in substantial part because its data were comparatively available to researchers. As a result, the empirical foundations of social media research have often reflected platform access as much as scientific design, raising longstanding concerns about platform-specific sampling frames, representativeness, and construct validity (boyd & Crawford, 2012; Hargittai, 2015; Olteanu et al., 2019).

That access has become increasingly unstable. Social media platforms have restricted, monetized, or discontinued research-facing APIs, leaving investigators dependent on access regimes that can change independently of the scientific value of the work they support (Freelon, 2018; Tromble, 2021; Davidson et al., 2023). These changes affect not only collection but methodological continuity: procedures built around discontinued or unequal access may be difficult to reproduce or independently validate, leaving an important part of the research infrastructure controlled by commercial entities outside the scientific process (Breuer et al., 2020).

Even when platform data are available, they have historically been dominated by public and platform-selected behavior. Public posts and interactions capture only one part of online life; direct messages, small-group communication, search and viewing histories, and media consumption can reflect interpersonal and behavioral processes that are not observable from public expression alone. This is especially consequential for questions about relationships, social support, harassment, well-being, or exposure, where the phenomenon may unfold in private interaction or in what people consume. Reliance on public data can therefore create an epistemic limitation: the portion of behavior made observable by a platform can become a proxy for the broader construct researchers intend to study (Howison et al., 2011; Cesare et al., 2018; Jungherr, 2018).

Participant-mediated data donation offers a complementary route to digital trace research. Rather than obtaining a platform-defined sample of user behavior, researchers can recruit individuals and invite them to contribute selected portions of downloadable archives made available through data-access rights (Ausloos & Veale, 2021; Ohme et al., 2024). This shifts the locus of access toward the participant and can make available longitudinal, non-public, cross-platform, and multimodal traces that are difficult to obtain through public APIs, while allowing participants to choose which platforms, data types, or periods of their history they contribute.

Participant-mediated access is more than an alternative mechanism for acquiring social media data. Because participants can contribute historical traces and link them to independently collected assessments, interviews, or dated events, donation can support retrospective, event-centered, and case-control designs in which the behavioral evidence is separated from the outcome against which it is evaluated (Stier et al., 2020; Parry et al., 2021).

These opportunities introduce methodological and ethical challenges. Platform archives are heterogeneous and mutable; successful donors may differ from non-donors; and archives may contain sensitive information about participants and third parties. Privacy strategies must also be construct-sensitive: some questions can be answered from minimized metadata, whereas others depend on language, images, audio, or conversational context. The methodological problem is therefore to identify the least intrusive representation that remains scientifically adequate and protect it throughout the data lifecycle.

Existing data donation infrastructures address important parts of this problem, including participant control, local processing, minimization, and linkage to study measures (Araujo et al., 2022; Boeschoten et al., 2023; Pfiffner et al., 2024; Kohne & Montag, 2024; Hakobyan et al., 2025). Research involving sensitive, semantically rich, multimodal, or cross-platform traces, however, can require additional capabilities after donation: platform-specific processing, protected persistent storage, study-level access governance, and adaptation as platform exports change. This motivates treating donation as an end-to-end research infrastructure rather than only an upload or extraction procedure.

In this paper, we present CANDOR (Collecting and Analyzing Networked Data for Open Research), an infrastructure for participant-mediated collection and governance of sensitive digital trace data. We derive design requirements for this class of research, describe how CANDOR operationalizes them, and compare its capabilities with existing data donation infrastructures. The comparison foregrounds tradeoffs among data minimization, construct preservation, study-design flexibility, and lifecycle governance. The generic portal, platform parsers and de-identification modules, data schemas, and deployment documentation will be released publicly upon acceptance.

## 2. Participant-Mediated Access to Digital Trace Data

### 2.1 From Platform-Mediated to Participant-Mediated Data Access

Platform-mediated access places consequential methodological decisions outside the research process: platforms determine which objects are available, at what granularity, over what period, and under what technical and financial conditions. As the “post-API age” has demonstrated, these conditions can change with little warning and can render established collection procedures unusable (Freelon, 2018; Tromble, 2021; Davidson et al., 2023). For longitudinal research programs, the result is a problem of methodological continuity as well as access.

Platform-mediated access also tends to privilege public and platform-selected traces. Ohme et al. (2024) distinguish these platform-centric approaches from user-centric collection, noting that participant-mediated donation can make non-public data such as private messages and individual profiling information available for research. This distinction matters because observable public content need not represent the broader behavioral process under study.

These access constraints also matter for cumulative science. Researchers may be unable to recreate a sampling frame, retrieve the same variables, or repeat an earlier collection procedure. Even when participant data cannot appropriately be made public, scientific scrutiny still depends on transparent and stable procedures for determining what was collected and transformed (Tufekci, 2014; Ruths & Pfeffer, 2014).

Participant-mediated data donation offers one such model. Under this approach, individuals obtain copies of digital trace data held about them and choose whether, and to what extent, to contribute those data to a research study. The approach has been enabled in substantial part by access and data-portability rights under the European Union’s General Data Protection Regulation (GDPR), through which platforms provide users with Data Download Packages (DDPs) containing portions of their historical activity (European Parliament & Council of the European Union, 2016; Boeschoten et al., 2022). Data donation consequently changes the locus of access: instead of a researcher requesting a platform-defined dataset about users, a participant exercises access to their own data and elects to contribute relevant portions for a specified research purpose.

Participant-mediated access should not be understood as replacing platform-mediated research access. Regulatory mechanisms can also provide independent researchers with non-public platform data; for example, the European Union’s Digital Services Act establishes a process for vetted researcher access to data from very large online platforms and search engines (European Parliament & Council of the European Union, 2022). Such mechanisms are important for platform accountability and population-level questions. Participant-mediated donation addresses a complementary need: it begins with an enrolled participant, can span multiple services used by that individual, and can connect historical traces to information collected directly from the same person.

## 2.2 Scientific Opportunities of Participant-Mediated Digital Trace Data

Retrospective depth enables observational designs that are difficult to implement with public social media data alone. Cases and controls can be defined using an independently measured clinical, behavioral, or survey criterion and comparable historical windows recovered for each group, as prior work linking participant-contributed social media histories to clinically ascertained outcomes illustrates (Birnbaum et al., 2019; Nguyen et al., 2022). Related designs can align within-person histories around events or repeated assessments. The key methodological advantage is that the outcome need not be defined from the same digital trace used as behavioral evidence. Figure 1 summarizes the research designs enabled by this combination of independently ascertained outcomes and retrospective longitudinal traces.

More generally, participant-mediated donation allows digital traces to be linked to surveys, interviews, psychological measures, clinical assessments, and other participant-level outcomes. Such linkage supports construct validation because digital traces are generated for purposes other than research and are not self-interpreting. Triangulation with independently measured constructs can reduce circular inference, a concern demonstrated in digital mental health research (Ernala et al., 2019; Chancellor & De Choudhury, 2020), while positioning traces as complementary rather than substitutive behavioral measures (Stier et al., 2020; Kmetty & Németh, 2022).

Data donation can also expand the range of observable behavior beyond public self-presentation. DDPs may include private messages, reactions, search histories, viewing or engagement records, media, and other platform-specific traces, depending on the service and the user's request. These data provide access to dimensions of digital life that are often unavailable through public APIs and scraping (Ohme et al., 2024). This distinction is particularly salient for interpersonal phenomena. A public post may reveal that an individual discusses loneliness or support, for example, but private communication can provide direct evidence about patterns of interaction, reciprocity, relationship dynamics, and the support exchanged between people. Similarly, research on online harms may require conversational context that is absent from isolated public posts, as participant-mediated studies of adolescents' Instagram data and Direct Messages have demonstrated (Razi et al., 2022, 2023; Ali et al., 2023). The value of non-public data therefore follows from the construct under study, rather than from an assumption that private data are inherently more informative.

Participant-mediated collection can also assemble selected traces from multiple services used by the same participant. This permits researchers to test whether signals are platform-specific or complementary and to characterize behavior across a broader digital ecology; cross-platform analyses of clinically grounded outcomes have demonstrated that such differences can matter empirically (Nguyen et al., 2022).

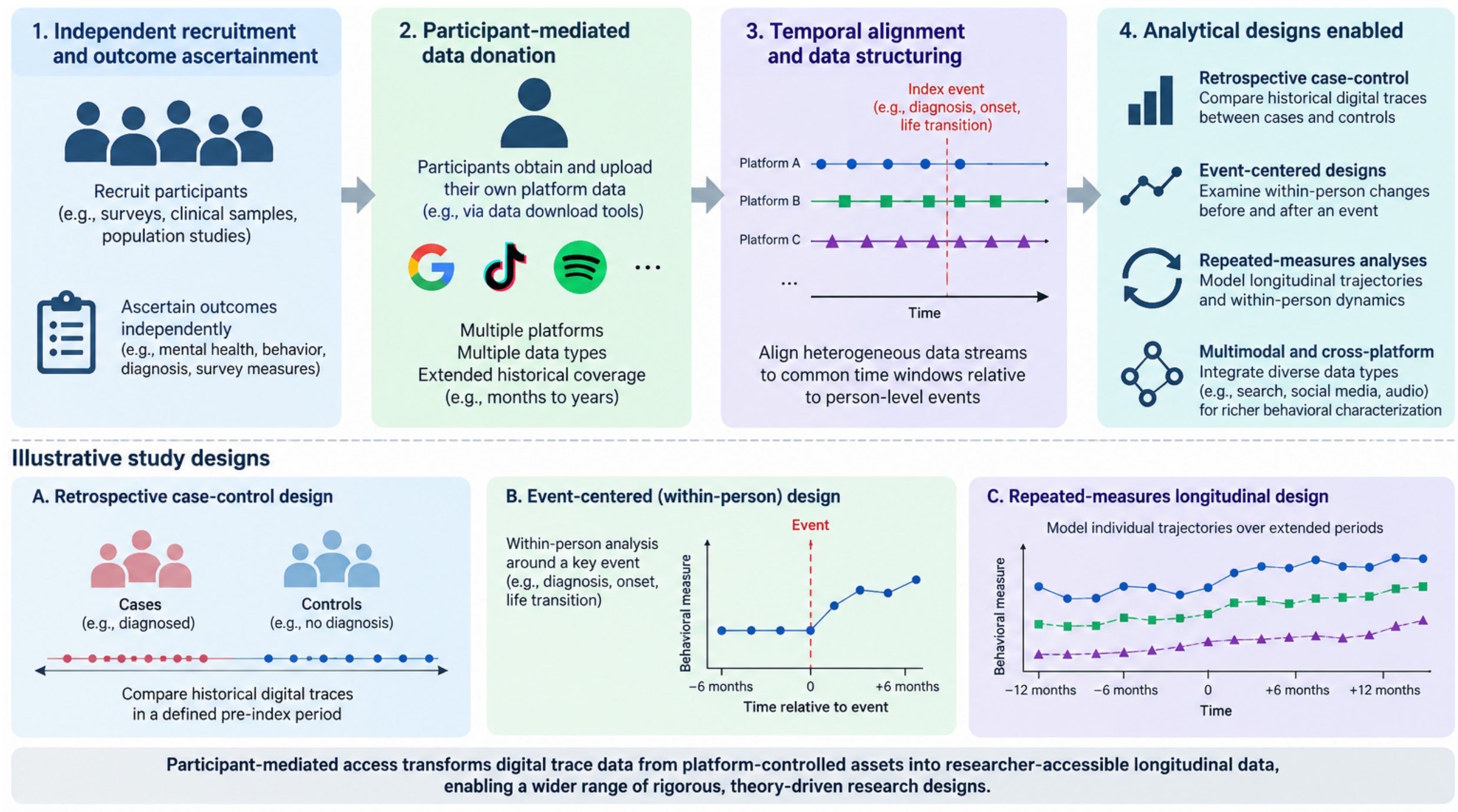

Figure 1. Research designs enabled by participant-mediated longitudinal digital trace data. Independent recruitment and outcome ascertainment can be linked to retrospective, multi-platform histories and aligned to common temporal windows, supporting case-control, event-centered, repeated-measures, and cross-platform analyses.

## 2.3 Methodological Challenges of Data Donation

Data donation does not eliminate the methodological limitations of digital trace research; it changes where several sources of error arise. The Total Error Framework for data donation emphasizes representation and measurement processes (Boeschoten et al., 2022a), extending broader total-error approaches for digital traces (Amaya et al., 2020; Sen et al., 2021). Carrière et al. (2025) translate these concerns into a study-design workflow spanning target population and data, recruitment, processing, donation, and analysis. A DDP should therefore not be treated as a neutral or complete behavioral record simply because it originates from a platform.

Selection into donation is one such concern, consistent with evidence that willingness to share passively collected or platform-derived data varies systematically across participants (Keusch et al., 2019; Struminskaya et al., 2020; Pfiffner & Friemel, 2023; Kmetty et al., 2025). Willingness to contribute digital data does not necessarily translate into successful donation, and the people who complete the process may differ from those who do not. In a study of more than 900 German Facebook users, Keusch et al. (2024) found that approximately 80% expressed willingness to donate, but only about one-third of those participants successfully completed the donation. Trust in researchers was positively associated with willingness and donation success. More recent cross-platform evidence also suggests that donation behavior and resulting nonresponse biases cannot be assumed to be constant across services (Wedel et al., 2026). In a study spanning YouTube, Facebook, Instagram, and TikTok, donation rates and predictors of donation differed by platform, with TikTok users less likely to complete donation than users of the other platforms studied. These findings make participation and attrition part of the inferential problem, particularly when a study seeks to compare behavior across platforms.

A second class of challenges concerns the contents of the DDP itself. Platform archives are created to satisfy user access requirements, not to serve as standardized research instruments, a limitation that has prompted calls for more stable and research-supportive platform data access and export mechanisms (Hase et al., 2024). The organization, naming, granularity, and completeness of exported data differ across platforms and can change over time (van Driel et al., 2022; Carrière et al., 2025). Researchers cannot assume that conceptually similar fields have equivalent meanings across services, or that an export obtained in one year will retain the same schema in a later study. These properties complicate reproducibility and place a maintenance burden on data donation infrastructure. Robust collection therefore requires platform-aware parsing and documentation, as well as procedures that can adapt when platforms alter their exports; related work on automatic DDP de-identification similarly highlights the need for platform-aware transformations (Boeschoten et al., 2021).

Privacy is a third and particularly consequential challenge. Donated archives may contain private conversations, embedded identifiers, media, and information about people who are not research participants. A participant's right to access an archive does not remove the obligation to minimize unnecessary exposure of either participants or third parties, particularly because expectations about appropriate information flows depend on context (Nissenbaum, 2004; Zimmer, 2010; Fiesler & Proferes, 2018). Data donation therefore requires explicit decisions about local processing, transfer, transformation, storage, access, and retention.

The principle of data minimization is essential, but its implementation depends on the scientific construct. In some studies, raw content is unnecessary. Dona, for example, was explicitly designed to study properties of social interaction while preventing raw messaging content from leaving the participant's browser. It retains de-identified metadata such as message timestamps and lengths, and its evaluation showed that these minimized traces can recover interactional properties including balance, heterogeneity, and burstiness (Hakobyan et al., 2025). This is a strong privacy-preserving design when the scientific question can be answered from those features. Other questions, however, depend on semantic, relational, or multimodal content itself. Distinguishing an expression of support from hostility, characterizing the meaning of a disclosure, or examining how a psychologically salient conversation evolves cannot generally be accomplished from message length and timing alone. In these settings, eliminating content may eliminate the construct being measured.

## 2.4 Existing Infrastructures for Data Donation

A growing set of tools has begun to operationalize participant-mediated data donation, alongside emerging methodological guidance emphasizing privacy protection, meaningful extraction, and user agency (Ohme & Araujo, 2022). One influential line of work emerged from OSD2F, an open-source data donation framework (Araujo et al., 2022), and the privacy-preserving data donation framework proposed by Boeschoten et al. (2022). The framework treats DDP donation as a participant-centered workflow in which only data relevant to a research question should be extracted and shared. Port subsequently implemented this approach as an open-source software tool for digital data donation (Boeschoten et al., 2023). Port processes DDPs locally on the participant's device, allowing researchers to specify extraction and transformation procedures while limiting the information transferred to the research team. This architecture is particularly valuable for minimizing unnecessary disclosure and for making the donation procedure configurable across studies.

The Data Donation Module (DDM) provides another general-purpose, open-source implementation. DDM allows researchers to configure data donation projects through a web application, define browser-side preprocessing and filtering rules, and link donated traces with self-reports collected through an integrated questionnaire or external survey software. It is designed for institutional deployment across multiple projects and emphasizes participant-facing usability, local processing, explicit consent, and encrypted storage (Pfiffner et al., 2024).

Other infrastructures have focused on particular forms of digital communication. ChatDashboard provides a framework for collecting, linking, and processing donated WhatsApp chat logs, combining an R package for parsing and anonymization with a web application through which participants can upload, review, and donate their chat histories (Kohne & Montag, 2024). Dona extends privacy-preserving donation to WhatsApp, Facebook, and Instagram messaging data and makes data minimization central to its design. Raw data are de-identified locally in the participant's browser; the donated representation contains anonymized identifiers and interaction metadata rather than message text. Dona also supports linkage to survey data through study-specific identifiers and provides researchers with an open-source deployment framework (Hakobyan et al., 2025). Related work has also demonstrated how donation workflows can be integrated with online survey systems to reduce fragmentation in participant-facing study procedures (Haim et al., 2023).

A broader class of behavioral and health research poses requirements that extend beyond the donation event. When semantic, multimodal, or cross-platform data constitute the phenomenon being measured, studies may need platform-specific processing, linkage to independently collected assessments, separation of research identifiers from personally identifying information, and continued protection through storage, access, retention, and deletion. These requirements motivate the design principles developed next.

## 3. Design Requirements for Participant-Mediated Research with Sensitive Digital Trace Data

### 3.1 Participant Agency, Transparency, and Proportional Data Collection

Access to a participant's digital archive does not imply that the entire archive is necessary for a study. The unit of consent should therefore be more granular than a binary decision to donate, consistent with consent-forward approaches emphasizing affirmative and legible choices (Pendse et al., 2024). Where exports permit it, participants should be able to choose platforms, data types, and temporal ranges, as prior Instagram data donation work has operationalized in a youth context (Razi et al., 2022). Researchers should correspondingly request the minimum information needed for the scientific aims rather than the maximum technically available.

Meaningful participant control also requires transparency about downstream use. Participants should be able to understand what requested files contain, how they will be processed, what analyses may be performed, who may access them, and whether raw or transformed versions will be retained. Because these implications may not be evident from platform-generated filenames or archive structures, donation interfaces should make data contents and downstream uses legible rather than treating consent as a one-time authorization for an opaque technical process (Fiesler & Proferes, 2018; Franzke et al., 2020; Pendse et al., 2024).

### 3.2 Scientific Grounding and Study Design Flexibility

Participant-contributed traces are most informative when interpreted alongside independently collected information about the participant and the phenomenon of interest. Infrastructure should therefore support linkage to study-specific surveys, interviews, psychological assessments, clinical information, or other outcomes while separating identifying information from analytic data. This enables researchers to test whether digital patterns correspond to constructs measured through established instruments rather than assuming that a behavioral trace is self-explanatory.

Because archives can contain behavior generated before enrollment, infrastructure should support retrospective alignment around independently ascertained outcomes or index events. In case-control designs, for example, historical traces can be aligned to comparable periods after cases and controls are defined using clinical, behavioral, or survey criteria. The infrastructure need not encode a particular inferential model, but it should preserve the linkage and temporal provenance required for such designs.

### 3.3 Privacy and Governance Across the Data Lifecycle

When the scientific question requires sensitive semantic or multimodal content, privacy protection cannot end with participant consent or a de-identification step. The relevant unit of protection is the full data lifecycle, consistent with lifecycle-oriented guidance for internet and big-data research ethics (Metcalf & Crawford, 2016; Franzke et al., 2020): receipt of the archive, transfer, temporary processing, de-identification, persistent storage, analytic access, retention, and deletion. Each stage creates different risks. Raw archives may contain direct identifiers; private conversations may contain information about communication partners who did not enroll in the study; images and videos can reveal faces or locations; and seemingly innocuous combinations of temporal and contextual information may permit re-identification. Infrastructure should therefore minimize the amount of time raw data remain exposed and constrain who and what systems can access them at each stage.

De-identification should likewise be treated as an ongoing technical process rather than a binary property of a dataset. Platform archives vary substantially in their content, and identifiers may appear in structured fields, free text, media, usernames, filenames, or platform-specific metadata. Effective processing may therefore require different approaches across data types and platforms. Automated procedures can reduce direct identifiers at scale, but no automated method can guarantee that all contextual or idiosyncratic identifiers will be detected; the broader re-identification literature cautions against treating removal of direct identifiers as equivalent to anonymity (Ohm, 2010). Infrastructure should make these limitations explicit and support modular de-identification procedures that can be revised as data formats and technical approaches change.

Sensitive digital trace research also requires governance after de-identification. Access should be limited by study and role, security-relevant events should be auditable, and retention should follow the governing protocol rather than default to indefinite storage. These controls are especially important when digital traces are linked to clinical or survey outcomes, because linkage increases analytic value while also increasing re-identification risk.

### 3.4 Adaptable and Reproducible Research Infrastructure

Participant-mediated research depends on an unstable upstream data source: commercial platform exports whose organization, fields, and available data types can change over time. Platform-specific parsers and transformations should therefore be modular and independently updateable, as should de-identification procedures across platforms and modalities.

Reproducibility in sensitive digital trace research should not be equated with unrestricted release of participant data. Private messages, health-related traces, and other participant-contributed archives may remain inappropriate for public dissemination even after automated de-identification (Zimmer, 2010; Williams et al., 2017). Reproducibility can instead be supported through reusable infrastructure, parsers, de-identification modules, schemas, deployment procedures, and documentation of study-specific configurations and transformations.

## 4. CANDOR: Design and Implementation

CANDOR was developed as an end-to-end infrastructure for participant-mediated collection of digital trace data, with particular attention to studies in which the data may be longitudinal, non-public, multimodal, or sensitive. The system separates participant-facing data receipt from protected research processing and storage, while allowing study teams to configure what data are requested and how donated archives are transformed. Its design reflects the requirements outlined above: participant-directed donation, proportional collection, linkage to independently collected study measures, platform- and modality-specific processing, and governance across the data lifecycle.

### 4.1 System Overview and Data Flow

CANDOR uses a two-tier architecture that separates the public-facing functions necessary for participant interaction from the protected environment in which donated data are processed and stored. The public-facing application provides study-specific instructions and receives participant uploads over encrypted connections. Donated files remain in this environment only temporarily. Following successful upload, files are transferred through an encrypted channel to a protected internal research environment that is not directly accessible from the public internet. Successful transfer is confirmed before the temporary copy is removed from the public-facing environment. This separation limits the period during which raw donated archives are present on an externally reachable system while retaining a participant-facing workflow that can be accessed without specialized institutional credentials.

CANDOR is provisioned at the study level. Before participant enrollment, an authorized administrator registers a study and creates a dedicated data structure for that project. Participants are subsequently registered using study-specific identifiers, and the system generates a unique upload link associated with the participant and study. Metadata needed to manage the donation and processing workflow are maintained separately by study, preventing identifiers or files from being inadvertently mixed across projects. The system does not require the participant's real-world identity to be encoded in donated filenames or analytic data. Identity crosswalks, where needed for recruitment or follow-up, remain under the governance of the corresponding study rather than serving as the primary identifier within CANDOR.

The workflow is platform-independent at the governance layer. Multiple platform archives can be associated with the same study-specific participant identifier while retaining platform, data-type, and temporal provenance. Within the protected environment, platform-specific modules transform relevant components into study-ready representations that can subsequently be linked to independently collected measures. Figure 2 shows both the conceptual data lifecycle and the underlying two-tier upload architecture.

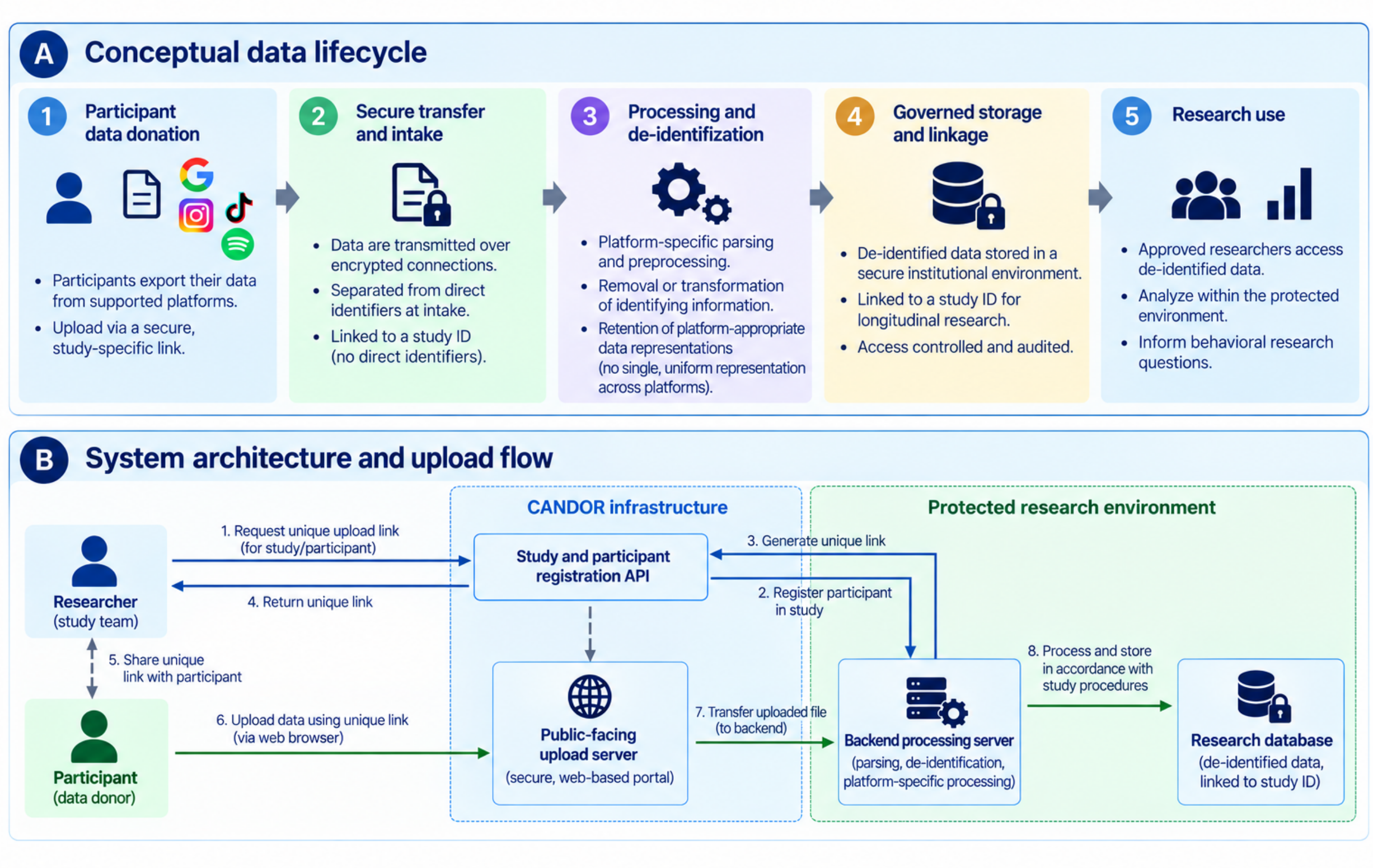

Figure 2. CANDOR architecture and system workflow. (A) Conceptual lifecycle separating participant-facing donation from protected processing, governed storage and linkage, and approved research use. (B) Two-tier upload architecture in which study teams generate participant-specific links, participants upload through a public-facing server, and archives are securely transferred to a protected backend for platform-specific processing and study-linked storage.

## 4.2 Participant-Facing Donation Workflow

Participants enter CANDOR through a study-specific workflow after completing the consent and eligibility procedures of the corresponding research protocol. The portal explains the purpose of data donation, the platforms supported by the study, the types of information that may be contained in platform exports, and the security practices governing upload and subsequent use. Platform-specific instructions guide participants through requesting and downloading their own archives. Because export procedures differ across services and change over time, these instructions are maintained separately for each platform rather than relying on a generic download procedure.

Donation is not treated as an all-or-nothing contribution. Participants can select supported platforms and, within them, the data types and temporal ranges they are willing to share. The resulting configuration is retained with the archive so downstream researchers can distinguish data that were not requested or selected from data absent for other reasons.

After receiving an archive from the platform, the participant uploads it using the unique study-specific link. During ingestion, files are assigned research identifiers that remove identifying information from filenames. Transfer to the protected research environment occurs over encrypted connections, and the system can notify designated study personnel when an upload and transfer have completed successfully. This workflow reduces the need for study teams to exchange sensitive archives through email, consumer cloud storage, or ad hoc file-transfer procedures.

The participant-facing interface also explains downstream analytic use, including examples of behavioral attributes or constructs that may be derived and the limits of such inference. This is particularly important for health or psychological research, where computational analysis could otherwise be mistaken for diagnosis or clinical assessment. Figure 3 shows representative views of the actual participant-facing CANDOR interface.

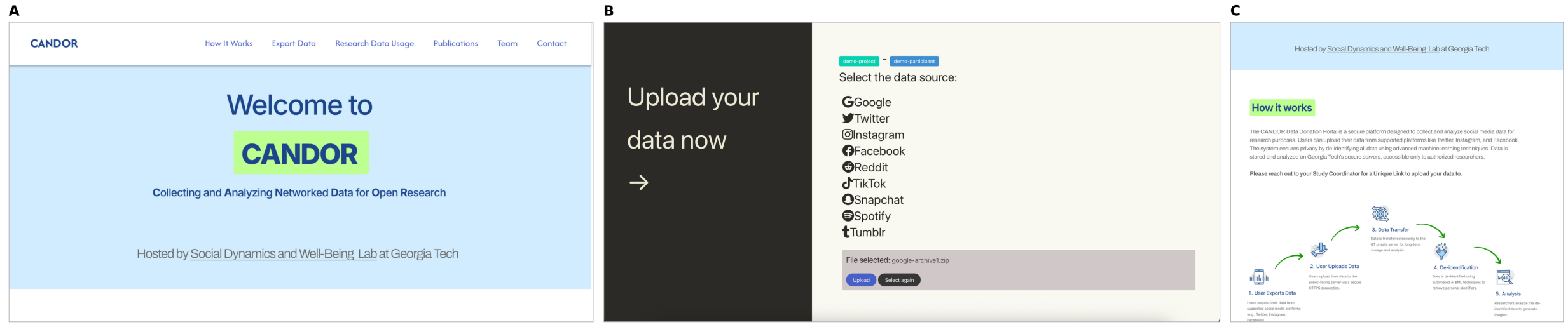

Figure 3. Participant-facing CANDOR interface. (A) CANDOR landing page and study explanation. (B) Study- and participant-specific platform selection and archive upload. (C) Participant-facing explanation of the data lifecycle, including transfer, de-identification, and analysis.

### 4.3 Platform-Specific Parsing, De-identification, and Multimodal Processing

Digital archives differ substantially in both structure and scientific content. CANDOR therefore separates the common donation and governance layer from platform-specific parsing and representation. Each supported platform has its own parser and associated processing modules, which can be updated without changing the broader upload, study-provisioning, or access-control infrastructure. The same modular principle is applied to de-identification. Platform and modality-specific procedures can be revised as export schemas change or stronger privacy-preserving methods become available. This design is important because a Google Search history, a TikTok watch history, and a Spotify listening history are not simply different file formats representing the same type of behavior; they capture qualitatively different forms of digital activity and require different transformations to become analytically useful.

For text-based archives, processing begins by normalizing platform-specific structures into study-ready records while retaining provenance and temporal information. In our Google Search pipeline, for example, exported search histories are converted into a common schema containing the study participant identifier, timestamp, query text, and relevant study information. Timestamps are normalized and temporal features are derived, while malformed records are excluded from temporal analyses. Search text is Unicode-normalized and processed using Presidio-based de-identification procedures (Microsoft, n.d.) that remove or replace potentially identifying elements such as URLs, email addresses, mentions, person and location entities, and numeric identifiers. The resulting records can be represented at multiple behavioral scales, including individual searches, temporally defined search sessions, days, and participant-level summaries. This permits the same underlying archive to support questions about individual information-seeking events as well as broader temporal patterns.

TikTok illustrates the additional processing required for multimodal consumption histories. Watch-history records identify videos encountered by the participant, but understanding the consumed content requires representations of the videos themselves. CANDOR's TikTok processing pipeline extracts three complementary modalities. Visual information is represented using transformer-based video embeddings; after evaluating VideoMAE (Tong et al., 2022) and X-CLIP (Ma et al., 2022), X-CLIP was selected as the primary representation model. Speech processing first uses YAMNet, a MobileNet-based AudioSet classifier (Google, n.d.), to distinguish speech-dominant, music-dominant, and mixed-audio content, followed by Whisper-small transcription for speech-dominant videos (Radford et al., 2023). Because TikTok videos frequently communicate through captions, stickers, and other text embedded in the visual stream, sampled frames are also processed using DeepSeek-OCR (Wei et al., 2025) to extract on-screen text. Video representations are passed through de-identification and dimensionality-reduction procedures intended to retain analytically salient information while reducing the ability to reconstruct the original visual content.

Spotify provides a different example in which the behavioral trace is a history of media consumption rather than user-authored content. CANDOR parses JSON-based listening histories containing playback timestamps, artist and track names, and playback duration. Unique tracks are identified within participant histories, both to characterize listening diversity and to avoid redundant downstream processing. Track records can then be enriched with acoustic attributes obtained through the ReccoBeats API (ReccoBeats, n.d.), including acousticness, danceability, energy, instrumentalness, liveness, loudness, speechiness, tempo, and valence. A shared cache stores previously retrieved track representations so that the same song does not require repeated external queries across participants. The resulting data retain both the temporal history of listening behavior and content-level acoustic attributes relevant to downstream behavioral analyses.

These examples illustrate a core distinction between platform ingestion and research representation. The platform determines the archive a participant receives; the scientific question determines which portions are retained and how they are represented. The modular processing layer allows heterogeneous traces to enter a common governed workflow without forcing them into a single representation. Figure 4 summarizes these platform-specific processing pathways and the resulting protected, study-linked representations.

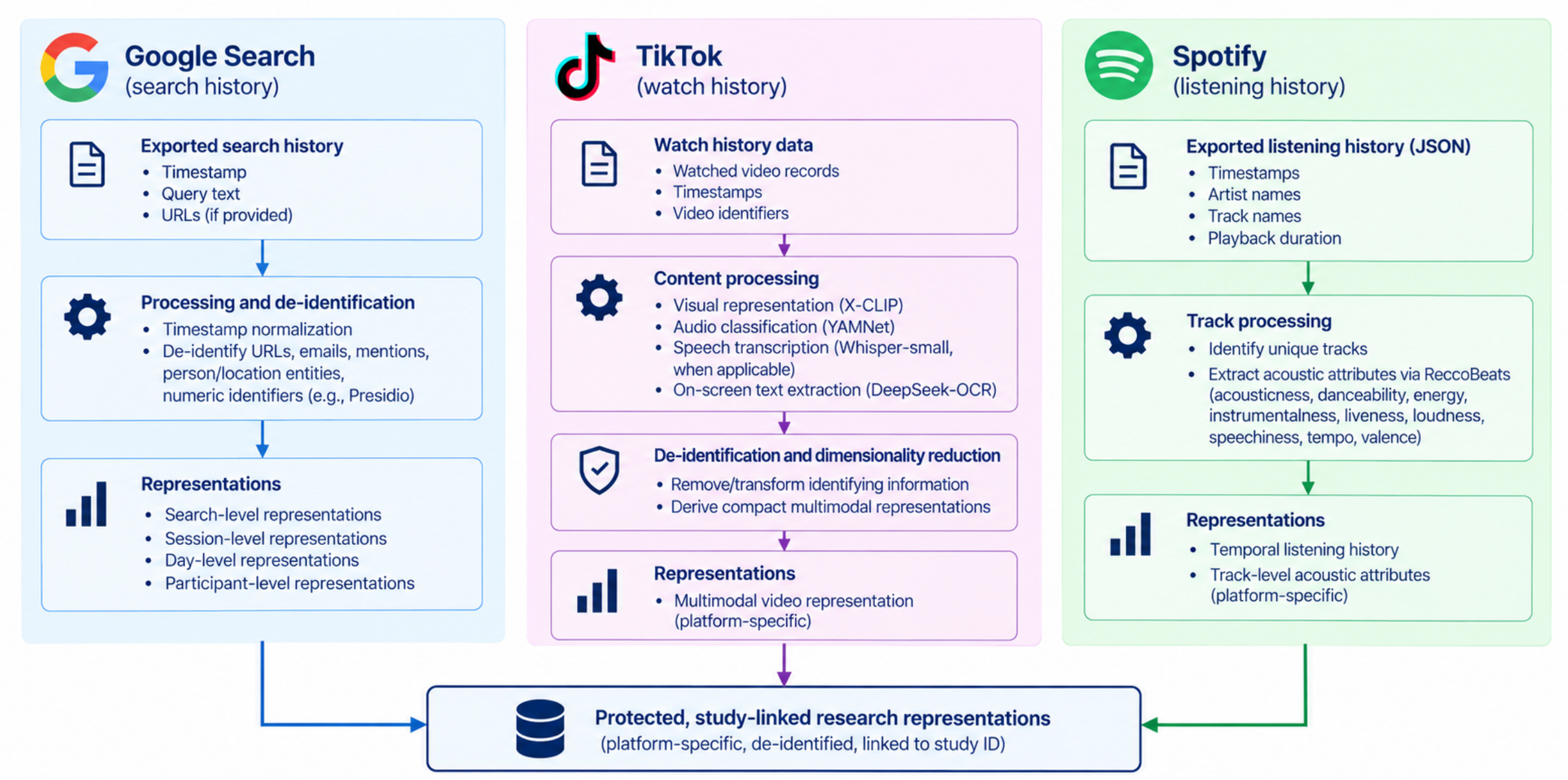

Figure 4. Platform-specific processing and research representations in CANDOR. Google Search histories are normalized and de-identified at multiple temporal granularities; TikTok watch histories are represented through visual, audio/speech, and on-screen-text processing; and Spotify listening histories are paired with track-level acoustic attributes. The resulting representations remain platform-specific while being protected and linked through study identifiers.

## 4.4 Study Linkage and Research-Ready Data

Each archive and processed representation remains associated with a study-specific participant identifier that authorized teams can use to link traces to separately collected surveys, interviews, psychological assessments, clinical measures, or other outcomes. Identifying crosswalks can remain under the more restrictive governance procedures of the parent study rather than appearing in analytic files.

Preserved timestamps, platform provenance, and participant linkage support retrospective and event-centered analyses, including alignment around hospitalization, diagnosis, exposure, or repeated assessments. CANDOR provides the data structures needed for these designs; their inferential assumptions remain properties of the individual research protocol.

Following platform-specific parsing and de-identification, research-ready data are organized by study and participant and stored in formats appropriate for analytic workloads. Structured outputs can be converted to columnar formats such as Parquet to support efficient retrieval and analysis of large longitudinal archives, while modality-specific representations can be retained in corresponding protected data structures. Retention of raw archives is governed by the study protocol: some protocols may permit temporary or longer-term retention under restricted access, whereas others may require deletion after de-identification and processing. CANDOR therefore separates the mechanics of ingestion and processing from study-specific decisions about what versions of the data may be retained.

### 4.5 Secure Storage, Access, and Governance

CANDOR's security architecture is designed around the sensitivity of the richest data it may receive. The protected research environment is accessible through institutional network controls and strong authentication; the public-facing application is limited to participant interaction and temporary upload. Data are encrypted in transmission and protected in storage, and the public-facing environment is not used as a long-term repository.

Access to research data follows the principle of least privilege (Saltzer & Schroeder, 1975). Personnel must be authorized for the relevant study and satisfy the human-subjects and data-use requirements associated with that protocol before access is provisioned. Study-level separation limits access to the projects for which an individual is approved. Researchers are not permitted to move protected data to personal devices, and access can be revoked when a person's role or study involvement ends. These controls are intended to ensure that the ability to administer or analyze one CANDOR study does not imply unrestricted access to data contributed to another.

Security and governance extend to infrastructure maintenance, including patching, vulnerability scanning, malware detection, file-integrity monitoring, authentication controls, security-event logging, and incident-response procedures. These operational controls matter because privacy protection depends on the computing environment throughout the period in which sensitive data are retained, not only on de-identification at ingestion.

CANDOR does not treat automated de-identification as a guarantee of anonymity. Naturalistic digital traces can contain nicknames, localized references, contextual identifiers, or combinations of information that automated systems may fail to recognize. The risk is amplified in private communication because archives can contain information about third parties as well as the participant. For this reason, de-identified data remain governed research data rather than being presumed safe for unrestricted release. Study teams can impose additional restrictions on raw or semantically rich content, and access to such material remains tied to the scientific need and approved protocol.

### 4.6 Reusability and Extensibility

Although developed within a specific institutional environment, CANDOR separates study configuration from the underlying donation infrastructure. Projects can specify platforms, data types, temporal ranges, processing procedures, and retention rules, while parsers and de-identification modules can be updated independently as exports or methods change.

The generic CANDOR portal, platform-specific parsers and de-identification modules, data schemas, and deployment documentation will be released publicly upon acceptance of this manuscript. The release will be designed to allow other research groups to instantiate and adapt the infrastructure within their own computing and data-governance environments. Institution-specific production configurations and security-sensitive operational details will not be included. This distinction reflects the broader approach to reproducibility described in Section 3: the procedures through which sensitive digital trace data are collected and transformed can be open and reusable even when the participant data themselves cannot appropriately be made public.

CANDOR therefore provides a governed pathway through which researchers can request proportionate data, preserve scientifically necessary information, link those traces to independently collected measures, and adapt processing as platforms and research questions change. The next section situates these capabilities within the broader data donation infrastructure landscape.

## 5. Comparative Capability Analysis

The systems reviewed in Section 2 were developed for overlapping but not identical research problems. We therefore compare them here not as competing implementations of a single ideal data donation architecture, but in terms of the research capabilities and privacy choices that their published descriptions support. The comparison focuses on OSD2F (Araujo et al., 2022), its successor line of development represented by Port and the current Port/Next data donation infrastructure (Boeschoten et al., 2023), DDM (Pfiffner et al., 2024), ChatDashboard (Kohne & Montag, 2024), Dona (Hakobyan et al., 2025), and CANDOR. Capabilities for the comparison systems were coded conservatively from their publications and current public documentation; an entry of "not established" indicates that we did not find sufficient documentation to characterize the capability, rather than evidence that the system cannot support it.

Table 1 shows that participant control and data minimization are common priorities across the existing infrastructures, but they are implemented differently. OSD2F allows participants to inspect the contents extracted from their Data Download Package and deselect individual items before donation (Araujo et al., 2022). Port extends this general approach through local extraction of researcher-specified features, participant review, and integration with a broader study environment; the current Port/Next infrastructure can support study setup, storage, survey integration, and study administration. DDM similarly performs researcher-defined preprocessing and filtering in the participant's browser before transmission and combines donation with integrated or externally linked survey measures (Pfiffner et al., 2024). Dona adopts a more restrictive representation for messaging data: raw message content is removed locally and only de-identified interaction metadata are donated. ChatDashboard is tailored to WhatsApp and supports upload, review, anonymization, linkage to survey responses, and retrospective analysis of communication histories. These approaches demonstrate that participant-mediated collection need not imply transfer of an entire raw archive to the researcher.

CANDOR shares the principle of proportional collection but occupies a different point in this design space. Participants can choose the platforms, data types, and temporal ranges they contribute, while the system can retain semantic or multimodal information when those data are required by the research question. This distinction is consequential for studies in which the construct is expressed in language, images, video, search behavior, or consumed media. CANDOR therefore does not use local elimination of raw content as a universal privacy strategy. Instead, it combines participant-directed scope with de-identification, secure transfer, protected processing and storage, study-specific access controls, and retention policies across the subsequent data lifecycle.

A second distinction concerns the breadth of the digital traces represented. ChatDashboard is intentionally specific to WhatsApp, and Dona focuses on messaging from WhatsApp, Facebook, and Instagram. OSD2F, Port/Next, and DDM are general-purpose frameworks whose study-specific extraction rules can be configured for different DDPs. CANDOR is similarly multi-platform, but its processing layer additionally treats platform and modality as analytically consequential. Platform-specific modules can transform Google Search histories into de-identified textual and temporal representations, TikTok watch histories into visual, speech, and on-screen-text representations, and Spotify histories into listening-event and acoustic-feature representations. Thus, multi-platform support refers not only to accepting heterogeneous archive formats, but to preserving the different behavioral constructs represented by those archives.

The infrastructures also differ in how far they extend beyond the donation event. OSD2F places much of post-collection security and data management on the deploying institution; Port/Next adds study storage, administration, and survey integration; and DDM provides encrypted storage while leaving downstream governance deployment-specific (Pfiffner et al., 2024). ChatDashboard and Dona provide deployable frameworks for their respective data types. CANDOR, by contrast, incorporates protected processing and persistent storage, study-level isolation and access provisioning, auditing, vulnerability management, and incident response as explicit parts of the deployment. These distinctions matter most when donated representations remain sensitive after preprocessing.

Linkage to independently measured outcomes is supported by several systems and is not unique to CANDOR. OSD2F can be embedded in survey flows; ChatDashboard and Dona support survey or questionnaire linkage; DDM includes integrated questionnaire functionality and external survey linkage; and Port/Next can integrate with survey platforms. CANDOR combines such linkage with preserved historical and temporal provenance to explicitly support retrospective case-control, event-centered, and longitudinal designs.

**Table 1. Capability comparison of participant-mediated digital trace infrastructures**

*Note. The table characterizes documented capabilities rather than assigning a performance score. "Deployment-specific" or "institution/deployment responsibility" indicates that the public materials describe the capability as dependent on how the research team hosts or configures the system. OSD2F is included because of its foundational role; Port represents a subsequent open-source implementation, and the current Port data donation task is designed to operate with the Next research platform. DDM refers to the Data Donation Module described by Pfiffner et al. (2024) and its current public documentation.*

## 5.1 Interpreting the Design Space

The comparison highlights two dimensions that are particularly important for behavioral research. The first is the relationship between data minimization and construct preservation. Dona demonstrates that highly minimized interaction metadata can be sufficient for questions about the temporal and structural properties of social interaction. This is preferable to transferring message content when content is unnecessary. In contrast, questions about the meaning of communication, psychological expression, exposure to particular content, or multimodal behavior may require representations that retain semantic information. The appropriate architecture therefore depends on what information is necessary to operationalize the construct, not on maximizing either data richness or data removal.

The second dimension is the scope of responsibility assumed by the infrastructure after a participant elects to donate. Local preprocessing can substantially reduce privacy risk before data leave the participant's device, but studies that require richer representations shift more responsibility to the receiving research environment. In such cases, secure transfer, de-identification, study isolation, access governance, auditing, retention, and incident response become part of the research method. CANDOR was designed around this end-to-end lifecycle because its intended use cases include data that remain sensitive even after automated de-identification.

These differences suggest that participant-mediated digital trace infrastructures should be evaluated in relation to the scientific and ethical requirements of the study rather than through a single ranking of capabilities. A narrowly tailored system can provide stronger minimization for a specific data type, while a general-purpose infrastructure can support a wider range of platforms and study designs at the cost of greater governance complexity. CANDOR is intended for the subset of studies in which longitudinal, linked, semantically rich, or multimodal traces are necessary and where those data therefore require continued protection within the research environment.

### 5.2 Reproducibility and Reuse

The comparison also distinguishes openness of infrastructure from openness of participant data. OSD2F, Port/Next, DDM, ChatDashboard, and Dona provide reusable implementations; CANDOR will similarly release its generic portal, parsers, de-identification modules, schemas, and deployment documentation. Across these systems, methodological procedures can therefore be inspectable and reusable even when study data cannot appropriately be released.

For sensitive participant-contributed data, this form of methodological transparency is often the more appropriate basis for reproducibility. Controlled-access derived data may additionally support secondary analysis where consent, governance, and residual disclosure risk permit it.

## 6. Discussion

The preceding design and comparison suggest that participant-mediated data donation is best understood as a research methodology rather than a single collection procedure. Its appropriate implementation depends on the construct being measured, the representation required to measure it, the study design in which the trace will be interpreted, and the risks that persist after donation. Five implications follow.

### 6.1 Data Minimization Should Be Defined Relative to the Scientific Construct

A central tension in data donation is between reducing the amount and sensitivity of information transferred to researchers and preserving what is necessary to study the phenomenon of interest. Framing this only as a privacy–utility tradeoff can obscure a prior methodological question: what representation of behavior is sufficient to operationalize the construct? If interaction frequency, timing, or reciprocity is the object of study, metadata may be sufficient and raw conversational content should not be collected. Dona demonstrates the value of this approach by extracting interactional properties while preventing message content from leaving the participant's device (Hakobyan et al., 2025).

For other constructs, however, semantic or multimodal information is not ancillary to the behavioral measure. Whether a message expresses support, coercion, distress, or humor depends on what is communicated and often on surrounding conversational context; participant-mediated studies of Instagram Direct Messages illustrate why relational context can be analytically necessary (Razi et al., 2023; Ali et al., 2023). Likewise, understanding consumed video may require visual, speech, and on-screen-text information. In such settings, aggressive removal of content can improve privacy while weakening construct validity. The methodological objective should therefore be to collect the least intrusive representation that remains adequate for the scientific question.

This principle places an additional obligation on researchers who determine that richer data are necessary. The justification for retaining semantic, relational, or multimodal information should be explicit, and protections after donation should increase with the sensitivity of the representation. Participant-directed selection, de-identification, restricted access, retention limits, and institutional security are complementary safeguards rather than substitutes for minimization. In this sense, the data request itself is part of research design: investigators should be able to explain why each requested platform, data type, and temporal window is necessary (Boeschoten et al., 2022; Carrière et al., 2025).

### 6.2 Participant-Mediated Data Can Expand Behavioral Study Design

The value of participant-mediated access extends beyond recovering data that are unavailable through APIs. Because participants can contribute historical archives after recruitment, digital traces can be paired retrospectively with independently measured outcomes. This creates a different inferential structure from studies in which both a psychological state and its behavioral correlates are inferred from the same observable content. Case status, symptoms, exposures, or other outcomes can instead be established through surveys, interviews, clinical assessments, or documented events and related to preceding digital behavior (Ernala et al., 2019; Stier et al., 2020).

This capability is particularly useful for case-control and event-centered designs. Researchers can identify cases and controls using an external criterion and recover comparable periods of activity preceding an index event; historical traces can likewise be aligned around hospitalization, diagnosis, life transitions, or repeated assessments. Prior work linking participant-contributed social media histories to clinically ascertained relapse and hospitalization illustrates this logic (Birnbaum et al., 2019; Nguyen et al., 2022). These designs do not eliminate confounding, selection, or temporal ambiguity, but they allow established principles of behavioral and epidemiological study design to be applied to naturalistic digital histories rather than requiring platform behavior itself to define the outcome.

Linkage also provides a route to stronger construct validation. Digital traces are behaviorally rich but were not generated as research measures; their interpretation benefits from triangulation with measures designed to assess the construct directly and, where possible, from participant accounts of what the observed behavior means in context (Kmetty & Németh, 2022; Parry et al., 2021). This is especially important in health and psychological research, where a computationally detectable pattern should not be treated as equivalent to a clinical state (Chancellor & De Choudhury, 2020). Participant-mediated collection can therefore position digital traces as complementary behavioral evidence within a multimethod study rather than as a replacement for established measurement.

### 6.3 Participant-Mediated Access Complements, Rather Than Replaces, Other Research Access Models

The contraction of public APIs has increased interest in data donation, but participant-mediated access is not a universal replacement for platform-mediated data (Freelon, 2018; Tromble, 2021; Davidson et al., 2023). Platform APIs and regulated researcher-access mechanisms remain better suited to questions about population-level diffusion, recommender systems, network structure, content moderation, or systemic platform effects. Participant-mediated access has a different comparative advantage: it begins with an enrolled individual, can combine traces from multiple services, can include non-public histories made available through personal exports, and can link those histories to information collected directly from the participant.

This distinction matters because the goal is not to reproduce API-based computational social science through another collection mechanism. Data donation enables a complementary form of digital behavioral research with different sampling frames, measures, strengths, and sources of error. It is especially well suited to questions in which individual-level context, validated outcomes, longitudinal history, or cross-platform behavior are central. Conversely, questions requiring population-level platform coverage or internal system data may be poorly served by participant donation alone.

The participant-centered sampling frame also requires careful attention to representativeness. Individuals who agree and successfully manage to donate may differ systematically from those who do not, and donation rates and predictors can vary across platforms (Keusch et al., 2024; Wedel et al., 2026). Meaningful participant choice introduces additional heterogeneity because participants in the same study may contribute different platforms, data types, or temporal ranges. Such variation is not simply a nuisance to be eliminated: it is partly a consequence of participant agency. But it must be recorded as provenance and incorporated into analysis rather than treated as undifferentiated missingness.

### 6.4 Open Science Does Not Require Open Sensitive Data

Participant-mediated research also exposes a tension in conventional discussions of open science. Raw or lightly de-identified archives may be inappropriate for public release when they contain private conversations, search histories, health-related behavior, or information about third parties. Automated de-identification cannot guarantee that contextual information will not permit re-identification (Ohm, 2010; Boeschoten et al., 2021), and republishing sensitive traces can create harms even when the original data were technically accessible (Zimmer, 2010; Williams et al., 2017).

For such data, the appropriate object of openness can instead be the research method and infrastructure. Collection software, platform parsers, de-identification procedures, schemas, analytic transformations, and deployment documentation can be made inspectable and reusable. Study-specific choices about requested data, temporal windows, processing, and retention can likewise be documented so that another team can reproduce the collection logic even when it cannot access the original archive. CANDOR's planned public release follows this model.

Controlled-access mechanisms may provide an additional route for secondary research when consent, protocol, and residual disclosure risk permit it. More generally, reproducibility for sensitive digital trace research should be evaluated in terms of whether provenance and transformation are transparent and whether the methodological pipeline can be independently inspected and instantiated—not solely by whether the underlying data can be posted publicly. This allows digital behavioral research to participate meaningfully in open science without treating participant privacy as an obstacle to be overcome.

### 6.5 Limitations and Future Directions

CANDOR does not resolve several limitations inherent to participant-mediated data donation. First, it depends on what platforms make available to users. DDPs differ in completeness and structure and may change without notice (van Driel et al., 2022; Carrière et al., 2025). Modular parsers reduce the cost of adapting to these changes but cannot recover information omitted from an export. A participant-contributed archive should therefore be treated as a platform-produced record with known provenance and coverage, not as a platform-independent ground truth.

Second, CANDOR reduces but does not eliminate privacy risk. Automated de-identification can miss contextual identifiers, local references, or information embedded in multimodal data, while private communications raise third-party privacy concerns not resolved by the donating participant's consent. For studies requiring such content, access restrictions and study-specific review remain necessary after automated processing. Future work should evaluate stronger privacy-preserving representations across modalities while measuring how much information can be removed without compromising the constructs for which the data were collected.

Third, participant choice and the procedural burden of obtaining archives can introduce selection and missingness (Carrière et al., 2025; Keusch et al., 2024; Wedel et al., 2026). Future deployments should document donation attempts, completion, and reasons for non-donation where protocols permit. Methodological work is also needed on how to model participant-selected platform, data-type, and temporal coverage, particularly when these choices correlate with the outcomes under study.

Finally, CANDOR is currently deployed within a particular institutional computing and governance environment. Although the generic portal, parsers, de-identification modules, schemas, and deployment documentation will be released upon acceptance, reproducing the full security and governance model will depend on local resources and requirements. Deployments across institutions with different technical and governance environments would help distinguish safeguards essential to sensitive participant-mediated research from those specific to one implementation.

Taken together, these limitations underscore that participant-mediated donation relocates rather than eliminates the methodological and ethical constraints of digital trace research. Its value lies in better aligning behavioral evidence with the scientific question while giving participants a more direct role in determining what is contributed; doing so requires collection, representation, linkage, privacy, and governance to be treated as parts of the same research method.

## 7. Conclusion

Participant-mediated data donation offers a complementary pathway for behavioral research as platform access contracts and consequential forms of digital behavior remain outside public data streams. Its methodological value depends on infrastructure that connects participant agency and proportional collection with scientifically appropriate representation, linkage to independent study measures, adaptable processing, and protection across the data lifecycle.

CANDOR provides one implementation of this approach. More broadly, the design space examined here suggests that data donation infrastructure should be matched to the scientific construct and study design: collecting no more information than necessary while preserving what is required to answer the research question. Aligning scientific necessity, participant control, and data governance is central to realizing the methodological promise of participant-mediated digital trace research.

## Author Contributions

Andrew Zhao: Software, Investigation. Rijul Magu: Methodology, Software. Ekta Raj: Software. Teresa Elinjikkal: Software. Munmun De Choudhury: Conceptualization, Writing – Original Draft, Supervision. All authors: Writing – Review & Editing.

## Open Practices Statement

The generic CANDOR portal, platform-specific parsers and de-identification modules, data schemas, and deployment documentation will be released publicly upon acceptance of this manuscript. Institution-specific production configurations and security-sensitive operational details will not be released. Participant-contributed digital trace data are not publicly available because they may contain sensitive, non-public, and third-party information.

## Ethics Approval

CANDOR is a research infrastructure that has been deployed within separately reviewed and approved human-subjects research protocols. This manuscript describes the design and implementation of the infrastructure and does not report a new human-subjects study.

## Consent to Participate

In studies using CANDOR, informed consent for participation and data donation is obtained in accordance with the procedures approved for the respective study protocol.

## Consent for Publication

Not applicable

## Competing Interests

De Choudhury serves on OpenAI's Expert Council on Well-Being and AI. The authors declare no other competing interests.

## Funding

This work was supported in part by the National Institute of Mental Health under awards P50MH115838, 5R01MH135488-02, and 1R01MH115905-01, and by a research agreement between the Georgia Institute of Technology and the University of Pittsburgh.

## Acknowledgments

We thank Rohan Garg, Chinar Dankhara, Trisha Jain, Gauri Sharma, and Irene Komal Paul Stephen for their contributions to the development of CANDOR and related research activities.

| Capability | OSD2F | Port / Next | ChatDashboard | Dona | DDM | CANDOR |
|---|---|---|---|---|---|---|
| General-purpose / multi-platform | Yes; configurable DDPs | Yes; current task supports multiple platforms | No; WhatsApp | Messaging: WhatsApp, Facebook, Instagram | Yes; configurable across DDPs | Yes; multiple social, search, media, and listening platforms |
| Participant inspection before donation | Yes | Yes | Yes / review workflow | Yes; anonymized data can be inspected | Yes; processed data presented before explicit consent | Transparency and scope selection; record-level inspection not currently a core feature |
| Granular participant control | Item-level (de)selection | Researcher extraction + participant consent/review | Chat selection / review | Time-window control; content minimized by design | Consent after researcher-defined preprocessing / filtering | Platform, data type, and time-range selection |
| Local / browser-side processing | Partial local extraction / selection | Yes | Anonymization within framework; deployment-specific | Yes; raw content remains on device | Yes; browser-side preprocessing and filtering | No; protected server-side processing |
| Semantic message/text content can be retained when required | Configurable | Configurable extraction | Yes; anonymized WhatsApp chat content can be processed | No; message text removed | Configurable; depends on extraction rules | Yes |
| Multimodal / non-text processing | Framework-dependent | Platform-script dependent | Limited to WhatsApp chat-log framework | Limited; voice-message metadata, media otherwise discarded | Platform-script / extraction-rule dependent | Yes; implemented text, video, speech/audio, on-screen text, and music/acoustic representations |
| Linkage to independent study measures | Yes; survey integration | Yes; e.g., Qualtrics integration | Yes; survey linkage | Yes; questionnaire linkage | Yes; integrated questionnaire or external survey linkage | Yes; surveys, interviews, clinical and other study measures |
| Retrospective temporal histories | Yes, where present in DDP | Yes, where present in DDP | Yes | Yes | Yes, where present in DDP | Yes |
| Supports case-control / event-centered designs | Possible through linkage; not a stated design focus | Possible through linkage; not a stated design focus | Possible through linkage | Possible through linkage | Possible through linkage; not a stated design focus | Explicitly supported through study identifiers, historical archives, and temporal alignment |
| Platform-specific modular parsers | Configurable extraction rules | Yes; per-platform extraction scripts | WhatsApp-specific parser (WhatsR) | Platform-specific messaging parsers | Configurable study-specific extraction rules | Yes |
| Modular de-identification / transformation | Preprocessing and minimization configurable | Configurable local transformations | WhatsApp anonymization/preprocessing | Fixed strong minimization for messaging | Yes; configurable preprocessing / filtering | Yes; platform- and modality-specific modules |
| Persistent study storage | Supported, deployment/institution dependent | Yes with Next | Yes, deployment-specific | Yes for minimized donated data | Yes; encrypted storage; deployment-specific | Yes; protected institutional environment |
| Study-level access governance / isolation | Institution/deployment responsibility | Study administration supported in Next | Deployment-specific | Deployment-specific | Deployment-specific; supports multiple institutional projects | Yes; study-specific provisioning and least-privilege access |
| Infrastructure auditing / security operations | Institution/deployment responsibility | Hosting/deployment dependent | Deployment-specific | Deployment-specific | Deployment / hosting dependent | Yes; institutional logging, vulnerability management, monitoring, incident |

| Capability | OSD2F | Port / Next | ChatDashboard | Dona | DDM | CANDOR |
|---|---|---|---|---|---|---|
| | | | | | | response |
| Open source / reusable implementation | Yes | Yes | Yes | Yes | Yes | Planned public release upon acceptance |

## References

Ali, S., Razi, A., Kim, S., Alsoubai, A., Ling, C., De Choudhury, M., Wisniewski, P. J., & Stringhini, G. (2023). Getting Meta: A multimodal approach for detecting unsafe conversations within Instagram Direct Messages of youth. Proceedings of the ACM on Human-Computer Interaction, 7(CSCW1), Article 132, 1–30. https://doi.org/10.1145/3579608

Amaya, A., Biemer, P. P., & Kinyon, D. (2020). Total error in a big data world: Adapting the TSE framework to big data. Journal of Survey Statistics and Methodology, 8(1), 89–119. https://doi.org/10.1093/jssam/smz056

Araujo, T., Ausloos, J., van Atteveldt, W., Loecherbach, F., Moeller, J., Ohme, J., Trilling, D., van de Velde, B., de Vreese, C., & Welbers, K. (2022). OSD2F: An open-source data donation framework. Computational Communication Research, 4(2), 372–387. https://doi.org/10.5117/CCR2022.2.001.ARAU

Ausloos, J., & Veale, M. (2021). Researching with data rights. Technology and Regulation, 2020, 136–157. https://doi.org/10.71265/sfcgjr17

Birnbaum, M. L., Ernala, S. K., Rizvi, A. F., Arenare, E., Van Meter, A. R., De Choudhury, M., & Kane, J. M. (2019). Detecting relapse in youth with psychotic disorders utilizing patient-generated and patient-contributed digital data from Facebook. npj Schizophrenia, 5, 17. https://doi.org/10.1038/s41537-019-0085-9

Boeschoten, L., Ausloos, J., Möller, J. E., Araujo, T., & Oberski, D. L. (2022a). A framework for privacy preserving digital trace data collection through data donation. Computational Communication Research, 4(2), 388–423. https://doi.org/10.5117/CCR2022.2.002.BOES

Boeschoten, L., de Schipper, N. C., Mendrik, A. M., van der Veen, E., Struminskaya, B., Janssen, H., & Araujo, T. (2023). Port: A software tool for digital data donation. Journal of Open Source Software, 8(90), 5596. https://doi.org/10.21105/joss.05596

Boeschoten, L., Mendrik, A., van der Veen, E., Vloothuis, J., Hu, H., Voorvaart, R., & Oberski, D. L. (2022b). Privacy-preserving local analysis of digital trace data: A proof-of-concept. Patterns, 3(3), 100444. https://doi.org/10.1016/j.patter.2022.100444

Boeschoten, L., Voorvaart, R., van den Goorbergh, R., Kaandorp, C., & De Vos, M. G. (2021). Automatic de-identification of Data Download Packages. Data Science, 4, 101–120. https://doi.org/10.3233/DS-210035

boyd, d., & Crawford, K. (2012). Critical questions for big data: Provocations for a cultural, technological, and scholarly phenomenon. Information, Communication & Society, 15(5), 662–679. https://doi.org/10.1080/1369118X.2012.678878

Breuer, J., Al Baghal, T., Sloan, L., Bishop, L., Kondyli, D., & Linardis, A. (2021). Informed consent for linking survey and social media data—Differences between platforms and data types. IASSIST Quarterly, 45(1). https://doi.org/10.29173/iq988

Breuer, J., Bishop, L., & Kinder-Kurlanda, K. (2020). The practical and ethical challenges in acquiring and sharing digital trace data: Negotiating public-private partnerships. New Media & Society, 22(11), 2058–2080. https://doi.org/10.1177/1461444820924622

Carrière, T. C., Boeschoten, L., Struminskaya, B., Janssen, H. L., de Schipper, N. C., & Araujo, T. (2025). Best practices for studies using digital data donation. Quality & Quantity, 59(Suppl 1), 389–412. https://doi.org/10.1007/s11135-024-01983-x

Cesare, N., Lee, H., McCormick, T., Spiro, E., & Zagheni, E. (2018). Promises and pitfalls of using digital traces for demographic research. Demography, 55(5), 1979–1999. https://doi.org/10.1007/s13524-018-0715-2

Chancellor, S., & De Choudhury, M. (2020). Methods in predictive techniques for mental health status on social media: A critical review. npj Digital Medicine, 3, 43. https://doi.org/10.1038/s41746-020-0233-7

Davidson, B. I., Wischerath, D., Racek, D., Parry, D. A., Godwin, E., Hinds, J., van der Linden, D., Roscoe, J. F., Ayravainen, L., & Cork, A. G. (2023). Platform-controlled social media APIs threaten open science. Nature Human Behaviour, 7, 2054–2057. https://doi.org/10.1038/s41562-023-01750-2

Ernala, S. K., Birnbaum, M. L., Candan, K. A., Rizvi, A. F., Sterling, W. A., Kane, J. M., & De Choudhury, M. (2019). Methodological gaps in predicting mental health states from social media: Triangulating diagnostic signals.

Proceedings of the 2019 CHI Conference on Human Factors in Computing Systems, Paper 134, 1–16. https://doi.org/10.1145/3290605.3300364

European Parliament & Council of the European Union. (2016). Regulation (EU) 2016/679 (General Data Protection Regulation). Official Journal of the European Union, L 119, 1–88.

European Parliament & Council of the European Union. (2022). Regulation (EU) 2022/2065 on a Single Market For Digital Services (Digital Services Act). Official Journal of the European Union, L 277, 1–102.

Fiesler, C., & Proferes, N. (2018). "Participant" perceptions of Twitter research ethics. Social Media + Society, 4(1). https://doi.org/10.1177/2056305118763366

Franzke, A. S., Bechmann, A., Ess, C. M., & Zimmer, M. (Eds.). (2020). Internet Research: Ethical Guidelines 3.0. Association of Internet Researchers.

Freelon, D. (2018). Computational research in the post-API age. Political Communication, 35(4), 665–668. https://doi.org/10.1080/10584609.2018.1477506

Google. (n.d.). YAMNet: Audio event classification. TensorFlow Hub. https://tfhub.dev/google/yamnet/1

Haim, M., Leiner, D., & Hase, V. (2023). Integrating data donations into online surveys. Medien & Kommunikationswissenschaft, 71(1–2), 130–137. https://doi.org/10.5771/1615-634X-2023-1-2-130

Hakobyan, O., Hillmann, P.-J., Martin, F., Böttinger, E., & Drimalla, H. (2025). Development and evaluation of Dona, a privacy-preserving donation platform for messaging data from WhatsApp, Facebook, and Instagram. Behavior Research Methods, 57, 94. https://doi.org/10.3758/s13428-024-02593-z

Hargittai, E. (2015). Is bigger always better? Potential biases of big data derived from social network sites. The ANNALS of the American Academy of Political and Social Science, 659(1), 63–76. https://doi.org/10.1177/0002716215570866

Hase, V., Ausloos, J., Boeschoten, L., Pfiffner, N., Janssen, H., Araujo, T., Carrière, T., de Vreese, C., Haßler, J., Loecherbach, F., Kmetty, Z., Möller, J., Ohme, J., Schmidbauer, E., Struminskaya, B., Trilling, D., Welbers, K., & Haim, M. (2024). Fulfilling data access obligations: How could (and should) platforms facilitate data donation studies? Internet Policy Review, 13(3), 1–37. https://doi.org/10.14763/2024.3.1793

Howison, J., Wiggins, A., & Crowston, K. (2011). Validity issues in the use of social network analysis with digital trace data. Journal of the Association for Information Systems, 12(12), 767–797. https://doi.org/10.17705/1jais.00282

Jungherr, A. (2018). Normalizing digital trace data. In N. J. Stroud & S. C. McGregor (Eds.), Digital discussions: How big data informs political communication (pp. 9–35). Routledge. https://doi.org/10.4324/9781351209434-2

Keusch, F., Pankowska, P. K., Cernat, A., & Bach, R. L. (2024). Do you have two minutes to talk about your data? Willingness to participate and nonparticipation bias in Facebook data donation. Field Methods, 36(4), 279–293. https://doi.org/10.1177/1525822X231225907

Keusch, F., Struminskaya, B., Antoun, C., Couper, M. P., & Kreuter, F. (2019). Willingness to participate in passive mobile data collection. Public Opinion Quarterly, 83(S1), 210–235. https://doi.org/10.1093/poq/nfz007

Kmetty, Z., & Németh, R. (2022). Which is your favorite music genre? A validity comparison of Facebook data and survey data. Bulletin of Sociological Methodology/Bulletin de Méthodologie Sociologique, 154(1), 82–104. https://doi.org/10.1177/07591063211061754

Kmetty, Z., Stefkovics, Á., Számely, J., Deng, D., Kellner, A., Pauló, E., Omodei, E., & Koltai, J. (2025). Determinants of willingness to donate data from social media platforms. Information, Communication & Society, 28(7), 1324–1349. https://doi.org/10.1080/1369118X.2024.2340995

Kohne, J., & Montag, C. (2024). ChatDashboard: A framework to collect, link, and process donated WhatsApp chat log data. Behavior Research Methods, 56, 3658–3684. https://doi.org/10.3758/s13428-023-02276-1

Lazer, D. M. J., Pentland, A., Watts, D. J., Aral, S., Athey, S., Contractor, N., Freelon, D., González-Bailón, S., King, G., Margetts, H., Nelson, A., Salganik, M. J., Strohmaier, M., Vespignani, A., & Wagner, C. (2020). Computational social science: Obstacles and opportunities. Science, 369(6507), 1060–1062. https://doi.org/10.1126/science.aaz8170

Lazer, D., & Radford, J. (2017). Data ex Machina: Introduction to big data. Annual Review of Sociology, 43, 19–39. https://doi.org/10.1146/annurev-soc-060116-053457

Lazer, D., Pentland, A., Adamic, L., Aral, S., Barabási, A.-L., Brewer, D., Christakis, N., Contractor, N., Fowler, J., Gutmann, M., Jebara, T., King, G., Macy, M., Roy, D., & Van Alstyne, M. (2009). Computational social science. Science, 323(5915), 721–723. https://doi.org/10.1126/science.1167742

Ma, Y., Xu, G., Sun, X., Yan, M., Zhang, J., & Ji, R. (2022). X-CLIP: End-to-End Multi-grained Contrastive Learning for Video-Text Retrieval. Proceedings of the 30th ACM International Conference on Multimedia, 638–647. https://doi.org/10.1145/3503161.3547910

Metcalf, J., & Crawford, K. (2016). Where are human subjects in big data research? The emerging ethics divide. Big Data & Society, 3(1). https://doi.org/10.1177/2053951716650211

Microsoft. (n.d.). Presidio: Data protection and de-identification SDK. https://microsoft.github.io/presidio/

Nguyen, V. C., Lu, N., Kane, J. M., Birnbaum, M. L., & De Choudhury, M. (2022). Cross-platform detection of psychiatric hospitalization via social media data: Comparison study. JMIR Mental Health, 9(12), e39747. https://doi.org/10.2196/39747

Nissenbaum, H. (2004). Privacy as contextual integrity. Washington Law Review, 79, 119–158.

Ohm, P. (2010). Broken promises of privacy: Responding to the surprising failure of anonymization. UCLA Law Review, 57, 1701–1777.

Ohme, J., & Araujo, T. (2022). Digital data donations: A quest for best practices. Patterns, 3(4), 100467. https://doi.org/10.1016/j.patter.2022.100467

Ohme, J., Araujo, T., Boeschoten, L., Freelon, D., Ram, N., Reeves, B. B., & Robinson, T. N. (2024). Digital trace data collection for social media effects research: APIs, data donation, and (screen) tracking. Communication Methods and Measures, 18(2), 124–141. https://doi.org/10.1080/19312458.2023.2181319

Olteanu, A., Castillo, C., Diaz, F., & Kıcıman, E. (2019). Social data: Biases, methodological pitfalls, and ethical boundaries. Frontiers in Big Data, 2, 13. https://doi.org/10.3389/fdata.2019.00013

Onnela, J.-P., & Rauch, S. L. (2016). Harnessing smartphone-based digital phenotyping to enhance behavioral and mental health. Neuropsychopharmacology, 41, 1691–1696. https://doi.org/10.1038/npp.2016.7

Parry, D. A., Davidson, B. I., Sewall, C. J. R., Fisher, J. T., Mieczkowski, H., & Quintana, D. S. (2021). A systematic review and meta-analysis of discrepancies between logged and self-reported digital media use. Nature Human Behaviour, 5, 1535–1547. https://doi.org/10.1038/s41562-021-01117-5

Pendse, S. R., Stapleton, L., Kumar, N., De Choudhury, M., & Chancellor, S. (2024). Advancing a consent-forward paradigm for digital mental health data. Nature Mental Health, 2, 1298–1307. https://doi.org/10.1038/s44220-024-00330-1

Pfiffner, N., & Friemel, T. N. (2023). Leveraging data donations for communication research: Exploring drivers behind the willingness to donate. Communication Methods and Measures, 17(3), 227–249. https://doi.org/10.1080/19312458.2023.2176474

Pfiffner, N., Witlox, P., & Friemel, T. N. (2024). Data Donation Module: A web application for collecting and enriching data donations. Computational Communication Research, 6(2), 1–19. https://doi.org/10.5117/CCR2024.2.4.PFIF

Radford, A., Kim, J. W., Xu, T., Brockman, G., McLeavey, C., & Sutskever, I. (2023). Robust speech recognition via large-scale weak supervision. Proceedings of the 40th International Conference on Machine Learning, 202, 28492–28518.

Razi, A., Alsoubai, A., Kim, S., Ali, S., Stringhini, G., De Choudhury, M., & Wisniewski, P. J. (2023). Sliding into My DMs: Detecting uncomfortable or unsafe sexual risk experiences within Instagram Direct Messages grounded in the perspective of youth. Proceedings of the ACM on Human-Computer Interaction, 7(CSCW1), Article 89, 1–29. https://doi.org/10.1145/3579522

Razi, A., Alsoubai, A., Kim, S., Naher, N., Ali, S., Stringhini, G., De Choudhury, M., & Wisniewski, P. J. (2022). Instagram Data Donation: A case study on collecting ecologically valid social media data for the purpose of adolescent online risk detection. Extended Abstracts of the 2022 CHI Conference on Human Factors in Computing Systems, Article 39, 1–9. https://doi.org/10.1145/3491101.3503569

ReccoBeats. (n.d.). ReccoBeats API documentation. https://reccobeats.com/docs/

Ruths, D., & Pfeffer, J. (2014). Social media for large studies of behavior. Science, 346(6213), 1063–1064. https://doi.org/10.1126/science.346.6213.1063

Salganik, M. J. (2018). Bit by Bit: Social Research in the Digital Age. Princeton University Press.

Saltzer, J. H., & Schroeder, M. D. (1975). The protection of information in computer systems. Proceedings of the IEEE, 63(9), 1278–1308. https://doi.org/10.1109/PROC.1975.9939

Sen, I., Flöck, F., Weller, K., Weiß, B., & Wagner, C. (2021). A total error framework for digital traces of human behavior on online platforms. Public Opinion Quarterly, 85(S1), 399–422. https://doi.org/10.1093/poq/nfab018

Silber, H., Breuer, J., Beuthner, C., Gummer, T., Keusch, F., Siegers, P., Stier, S., & Weiß, B. (2022). Linking surveys and digital trace data: Insights from two studies on determinants of data sharing behaviour. Journal of the Royal Statistical Society: Series A (Statistics in Society), 185(Suppl. 2), S387–S407. https://doi.org/10.1111/rssa.12954

Stier, S., Breuer, J., Siegers, P., & Thorson, K. (2020). Integrating survey data and digital trace data: Key issues in developing an emerging field. Social Science Computer Review, 38(5), 503–516. https://doi.org/10.1177/0894439319843669

Struminskaya, B., Toepoel, V., Lugtig, P., Haan, M., Luiten, A., & Schouten, B. (2020). Understanding willingness to share smartphone-sensor data. Public Opinion Quarterly, 84(3), 725–759. https://doi.org/10.1093/poq/nfaa044

Tong, Z., Song, Y., Wang, J., & Wang, L. (2022). VideoMAE: Masked autoencoders are data-efficient learners for self-supervised video pre-training. Advances in Neural Information Processing Systems, 35, 10078–10093.

Tromble, R. (2021). Where have all the data gone? A critical reflection on academic digital research in the post-API age. Social Media + Society, 7(1). https://doi.org/10.1177/2056305121988929

Tufekci, Z. (2014). Big questions for social media big data: Representativeness, validity and other methodological pitfalls. Proceedings of the International AAAI Conference on Web and Social Media, 8(1), 505–514. https://doi.org/10.1609/icwsm.v8i1.14517

van Driel, I. I., Giachanou, A., Pouwels, J. L., Boeschoten, L., Beyens, I., & Valkenburg, P. M. (2022). Promises and pitfalls of social media data donations. Communication Methods and Measures, 16(4), 266–282. https://doi.org/10.1080/19312458.2022.2109608

Wedel, L., Ohme, J., Mayer, A.-T., Gaisbauer, F., & Fan, Y. (2026). The platform matters: Cross-platform differences in data donation willingness, behavior, and bias. Communication Methods and Measures, 20(1), 53–77. https://doi.org/10.1080/19312458.2025.2605946

Wei, H., Sun, Y., & Li, Y. (2025). DeepSeek-OCR: Contexts optical compression. arXiv. https://doi.org/10.48550/arXiv.2510.18234

Williams, M. L., Burnap, P., & Sloan, L. (2017). Towards an ethical framework for publishing Twitter data in social research: Taking into account users' views, online context and algorithmic estimation. Sociology, 51(6), 1149–1168. https://doi.org/10.1177/0038038517708140

Zimmer, M. (2010). "But the data is already public": On the ethics of research in Facebook. Ethics and Information Technology, 12, 313–325. https://doi.org/10.1007/s10676-010-9227-5